\documentclass[onecolumn,nofootinbib]{revtex4}
\usepackage{amsmath}
\usepackage{graphicx}
\usepackage{amssymb}
\usepackage{float}
\usepackage{mathpazo}
\usepackage{lineno}
\usepackage{subfigure}

\newcommand{\ignore}[1]{}

\newcommand{\be}{\begin{equation}}
\newcommand{\ee}{\end{equation}}
\def \ba#1\ea{\begin{align}#1\end{align}}
\newcommand{\bit}{\begin{itemize}}
\newcommand{\eit}{\end{itemize}}

\def \slashb#1{\setbox0=\hbox{$#1$}#1\hskip-\wd0\dimen0=5pt\advance
        \dimen0 by-\ht0\advance \dimen0 by\dp0\lower0.5\dimen0\hbox
          to\wd0{\hss \sl/\/ \hss}}

\input{epsf}
\input{tcilatex}
\begin{document}

\title{Sensitivities to charged-current nonstandard neutrino interactions at
DUNE }
\author{\textbf{Pouya Bakhti}}
\affiliation{School of Physics, Institute for research in fundamental sciences (IPM), PO
Box 19395-5531, Tehran, Iran}
\author{\textbf{Amir N. Khan}}
\affiliation{School of Physics, Sun Yat-Sen University (SYSU), Guangzhou, 510275, China}
\author{\textbf{W. Wang}}
\affiliation{School of Physics, Sun Yat-Sen University (SYSU), Guangzhou, 510275, China}
\email{pouya\_bakhti@ipm.ir, khan8@mail.sysu.edu.cn,
wangw223@mail.sysu.edu.cn }

\begin{abstract}
We investigate the effects of charged-current (CC) nonstandard neutrino
interactions (NSIs) at the source and at the detector in the simulated data
for the planned Deep Underground Neutrino Experiment (DUNE), while
neglecting the neutral-current NSIs at the propagation due to the fact that
several solutions have been proposed to resolve the degeneracies posed by
neutral-current NSIs while no solution exists for the degeneracies due to
the CC NSIs. We study the effects of CC NSIs on the simultaneous
measurements of $\theta _{23}$ and $\delta _{CP}$ in DUNE. The analysis
reveals that 3$\sigma $ C.L. measurement of the correct octant of $\theta
_{23}$ in the standard mixing scenario is spoiled if the CC NSIs are taken
into account. Likewise, the CC NSIs can deteriorate the uncertainty of the $%
\delta _{CP}$ measurement by a factor of two relative to that in the
standard oscillation scenario. We also show that the source and the detector
CC NSIs can induce a significant amount of fake CP-violation and the
CP-conserving case can be excluded by more than 80\% C.L. in the presence of
fake CP-violation. We further find the potential of DUNE to constrain the
relevant CC NSI parameters from the single parameter fits for both neutrino
and antineutrino appearance and disappearance channels at both the near and
far detectors. The results show that there could be improvement in the
current bounds by at least one order of magnitude at the near and far
detector of DUNE except a few parameters which remain weaker at the far
detector.\textbf{\ }
\end{abstract}

\maketitle

\section{Introduction}

The discovery of the nonzero value of the neutrino mixing angle $\theta
_{13} $ has revolutionized the field of neutrino physics in the recent
years. This has been made possible after the twenty years long efforts to
build and improve the sophisticated detector technology for the measurement
of the small value of $\theta _{13}$ in the accelerator neutrino experiment,
T2K \cite{Abe:2011sj} and the reactor neutrino experiments, Double Chooz 
\cite{DC}, RENO \cite{RENO} and Daya Bay \cite{DB}. These experiments have
measured the value of $\theta _{13}$ with unprecedented precision. The next
goals in the neutrino oscillation study is to find the correct ordering of
the neutrino mass hierarchy (Normal or Inverted), the determination of the
CP-violating phase ($\delta _{CP}$), to find the correct octant of the
mixing parameter $\theta _{23}$ and the precise measurements of all the
parameters of the neutrino oscillation scheme. All these unknown parameters
and the information will be explored in the medium-baseline reactor neutrino
oscillation experiments, JUNO \cite{JUNO} and RENO-50 \cite{RENO-50}, and in
the long baseline accelerator experiments, T2K \cite{Abe:2014tzr}, NO$\nu $A 
\cite{Adamson:2016xxw}, T2HK \cite{Abe:2015zbg} and the Deep Underground
Neutrino Experiment (DUNE) \cite{DUNE}.

Two main goals out of the several others in the ongoing DUNE are the
measurement of $\delta _{CP}$ and determination of the correct octant of $%
\theta _{23}$. The apparent issue in the simultaneous measurement of these
two parameters is their strong correlation with each other. Previously, it
has been shown that in $\nu _{e}/\overline{\nu }_{e}$ appearance channels,
the simultaneous measurement of $\theta _{23}$ and $\delta _{CP}$ in the
long baseline experiments in the region 40$^{0}\leq $ $\theta _{23}\leq $ 50$%
^{0}$ is the better way to measure the two parameters more precisely \cite%
{Minakata:2013hgk}. The same authors have studied the correlation and their
impacts on the measurement of the two parameters by considering both the
appearance and disappearance channels at the long baseline experiments, T2HK 
\cite{t2khk}, LBNE \cite{LBNE}, IDS-NF \cite{idsf} and ESS$\nu $SB \cite%
{essnu} with different baselines \cite{Coloma:2014kca}. It has been proven
that how the interplay between the two channels can improve the measurements
of $\theta _{23}$ and $\delta _{CP}$ and how the existing degeneracy can be
lifted.

In the recent years, after the discovery of the nonzero value of $\theta
_{13}$, a wide effort has been put to investigate the hints for NSIs at the
neutrino sources and detectors in the reactor short-baseline experiments 
\cite{ANK1,ANK2,ANK3,Girardi:2014kca,Girardi:2014gna} and for the NSIs at
propagation in the accelerator long baseline experiments \cite%
{deGouvea:2015ndi,Liao:2016hsa,Coloma:2015kiu,Coloma:2016gei,Masud:2016gcl}.
In the latter case, it has been demonstrated recently in Ref. \cite%
{deGouvea:2015ndi} that how the data from DUNE setup can be modified in the
presence of different scenarios of NSIs at propagation of magnitude larger
than $O(10^{-1}$) if the DUNE data is not consistent with the standard
paradigm. There exists degeneracy between the standard mixing parameters and
NSI parameters at propagation of the appearance channels in the three long
baseline neutrino oscillation experiments T2K, NOVA and DUNE \cite%
{Liao:2016hsa} . It was found that at a single $L/E_{\nu }$, both diagonal
and off-diagonal NSI parameters can lead to the four-fold degeneracy that
can affect the measurements of mass hierarchy, octant of $\theta _{23}$ and $%
\delta _{CP}$. It was also shown that the degeneracy cannot be resolved even
by the combined data of T2K and NO$\nu $A, however a wide-band beam
experiment like DUNE can resolve the degeneracy in some scenarios, but not
for the others. It was further shown in Ref. \cite{Bakhti:2016prn}, that
such a degeneracy can also be resolved with the help of a low-energy
neutrino oscillation experiment such as MOMENT \cite{Cao:2014bea}.\textbf{\ }%
Some other recent work related to the new physics sensitivities at DUNE can
be found in Ref. \cite{ska1,bkays,ska2,valle}\textbf{.}

In this work, we consider an alternative approach by considering the
charged-current (CC) NSIs which can affect the neutrino production and
detection for the case of DUNE. Although neutral-current (NC) NSI which are
active during the propagation through the earth matter are more important
for the high energy neutrino beams of DUNE, however, the CC NSI at the
source and detector are also equally important and cannot be ignored.\textbf{%
\ }Since the standard and non-standard NC interactions at the propagation in
matter depend on the neutrino energy \cite{Ohlsson:2012kf}, therefore by
combining DUNE, which is a high energy neutrino experiment, with another low
energy neutrino experiment such as MOMENT, the degeneracies posed by the NC
NSIs to the determination of oscillation parameters can be resolved.
However, if the degeneracies produced are due to the CC NSIs, then they
cannot be resolved in this way, (see Ref. \cite{Bakhti:2016prn} for
details). Moreover, for the near detector only the CC NSIs are relevant
while the NC NSIs play no role at the near detector of DUNE, therefore we
include the near detector simulated data in addition to far detector. From
the near detector alone, we obtain stronger bounds than the existing bounds
on the CC NSI parameters as given in the 3rd column of Table I for the 3+3
years of DUNE run. It is also important to note that in future if the DUNE
near detector data is combined with the other short-baseline low energy
experiments, they will constrain the CC NSI parameters up to several orders
of magnitude.

In this work, first we analyze the low-level information at the probability
and at the event rate spectrum level and then perform the full statistical
analysis to find the sensitivity of DUNE measurements to the CC NSIs at the
source and at the detector. We explore the correlation between the two
standard parameters $\theta _{23}$ and $\delta _{CP}$ in the presence of CC
NSIs. Analysis with the same goals, but for the different baseline at the ESS%
$\nu $SB facility \cite{essnu} has been carried out in Ref. \cite%
{Blennow:2015nxa}, where it has been shown that the precision of $\theta
_{23}$ is robust in the presence of source and detector NSIs, but $\delta
_{CP}$ measurement gets worsen. Ref. \cite{Blennow:2016etl}\footnote{%
This paper was submitted to arXiv one day before our work.} also studies the
effects of NC and CC NSI at DUNE in a different perspective, where they
consider only the far detector in their analysis and follow an unrealistic
approach by considering different NSI at the source and detector. Contrary
to Ref. \cite{Blennow:2016etl}, we use the simulated data from both the near
and far detectors with equal emphasis on both and taking the advantage of
the fact that the NSI at production and detection are the same. We find the
effects of the CP-violation caused by NSI phases when the standard $\delta
_{CP}=0$; we called it the fake CP-violation and denote it by $\delta
_{CP}^{\prime }$. Further, we explore the potential of DUNE to constrain the
CC NSI parameters at the neutrino source and at the near and far detector
and compare them with the current bounds.

This paper is organized as follows. In section \ref{theo}, we present the CC
effective four-Fermion NSI Lagrangian and discuss the effects of NSI on
oscillation probability. In section \ref{anal}, we discuss about the
characteristics of DUNE. The details of our analysis and the results are
presented in section \ref{result}. The summary and conclusions to this work
are given in section \ref{sum}.

\section{Theoretical Framework\label{theo}}

For the DUNE\ set up, high energy beams (up to 10GeV) of neutrino and
antineutrinos are produced at the accelerator as a result of pion decays $%
(\pi ^{-}\rightarrow \mu ^{^{-}}+\overline{\nu }_{\alpha }\ $ and $\pi
^{+}\rightarrow \mu ^{^{+}}+\nu _{\alpha })$. 
They travel over a distances of 1300 km towards the\textbf{\ }DUNE far
detector and over a distance of 450m towards the DUNE near detector and
their detection is made through inverse beta decays ($\nu _{\alpha
}n\rightarrow pe^{-}$ and $\overline{\nu }_{\alpha }p\rightarrow ne^{+}$) in
the liquid argon scintillation detector. 
The left-handed four-Fermion CC effective NSI Lagrangian which governs the
pion decays at the source and the inverse beta decays at the detectors are
given by \cite{ANK1,ANK2,ANK3} 
\begin{equation}
\mathcal{L}_{SM+NSIs}=-2\sqrt{2}G_{F}(\delta _{\alpha \beta }+\varepsilon
_{\alpha \beta }^{udL})(\overline{l_{\alpha }}\gamma _{\lambda
}P_{L}U_{\beta a}\nu _{a})(\bar{d}\gamma ^{\lambda }P_{L}u)^{\dagger }+h.c.,
\label{eq.1}
\end{equation}%
%
%
%
%
%
%
%
%
%
%
%
%
%
%
%
%
%
%
%
%
%
%
%
%
%
%
%
%
%
%
%
%
%
%
%
%
%
where $\alpha ,\beta $ are the flavor indices and $a$ is the mass index and
all the repeated indicies are summed over. We have restricted ourselves to
the case of left-handed neutrino helicity and vector and axial-vector quark
currents. For simplicity, we do not consider any right-handed NSI couplings.
The coefficients $\varepsilon _{\alpha \beta }^{udL},\ $which are complex in
general, are the relative coupling strengths of the different flavor
combinations of NSI to the standard model semileptonic, 
while in the SM case $\varepsilon _{\alpha \beta }^{udL}=0$. 
Here U is the standard leptonic mixing matrix, parametrized in the standard
form. For brevity, we denote all NSI parameters ($\varepsilon _{\alpha \beta
}^{udL}$) at source and detector by $\varepsilon _{\alpha \beta }^{s}\ $and $%
\varepsilon _{\alpha \beta }^{d}$\ where the indices 's' and 'd' stand for
the 'source' and 'detector'.\textit{\ }Since the production and detection of
the neutrinos have the same interaction process at the quark level, we can
take $\varepsilon _{\alpha \beta }^{s}=\varepsilon _{\alpha \beta }^{d^{\ast
}}$\ for the neutrino and for $\varepsilon _{\alpha \beta }^{d}=\varepsilon
_{\alpha \beta }^{s^{\ast }}$\ for the antineutrino beams and thus remove
"s" and "d" indices further and denote all the parameters by $\varepsilon
_{\alpha \beta }$ in general. This pragmatic choice reduces the number of
NSI parameters to a half and as a result give better parameter fits. This
fact has not been considered in ref. \cite{Blennow:2016etl} and therefore
they get weaker constraints as compared ours on the NSI parameters.\ Here
for $\alpha =\beta $\ in the Eq. (1), the summation corresponds to the SM
and non-universal flavor diagonal NSI while $\alpha \neq \beta $\
corresponds to the flavor-violating NSI.

\begin{table*}[tbph]
\begin{center}
\begin{tabular}{|c|c|c|c|}
\hline
Channel & standard parameters (FD) & CC-NSI parameters (FD) & CC-NSI
parameters (ND) \\ \hline
$P(\nu _{\mu }(\overline{\nu }_{\mu })\longrightarrow \nu _{e}(\overline{\nu 
}_{e}))$ & $\theta _{23}$, $\theta _{12}$, $\theta _{13}$, $\Delta
m_{31}^{2} $, $\Delta m_{21}^{2}$, $\delta _{CP}$ & $\varepsilon _{\mu e}$, $%
\varepsilon _{\tau e}$ & $\varepsilon _{\mu e}$ \\ \hline
$P(\nu _{\mu }(\overline{\nu }_{\mu })\longrightarrow \nu _{\mu }(\overline{%
\nu }_{\mu }))$ & $\theta _{23}$, $\Delta m_{31}^{2}$ & $\varepsilon _{\mu
\mu }$, $\varepsilon _{\mu \tau }$, $\varepsilon _{\tau \mu }$, $\varepsilon
_{e\mu }$ & $\varepsilon _{\mu \mu }$ \\ \hline
$P(\nu _{e}(\overline{\nu }_{e})\longrightarrow \nu _{e}(\overline{\nu }%
_{e}))$ & $-$ & $-$ & $\varepsilon _{ee}$ \\ \hline
\end{tabular}%
\end{center}
\caption{Standard oscillation and NSI parameters that the near and far
detectors are sensitive to them for appearance and disappearance channels.
Notice the near detector is not sensitive to the standard oscillation
parameters. }
\label{tab:osc}
\end{table*}

\ In our simulation we used GLoBES, and for including the NSI we used
oscillation probability formulas from Ref. \cite{Kopp:2007ne,Kopp:2006wp},
so we do not repeat the lengthy analytical probability formulas which were
calculated using perturbative expansions upto the leading order in the NSI
parameters, however we summarize the probabilities and their NSI parameters
dependences in table \ref{tab:osc}\textit{.} In case of the near detector,
the baseline L=0, therefore the oscillation probability at the near detector
does not depend on standard oscillation parameters, and for electron
appearance, it depends on $\varepsilon _{\mu e}$\ and for disappearance
mode, it depends on $\varepsilon _{\mu \mu }~\ $and $\varepsilon _{ee}$\ as
shown in TABLE\  \textit{\ref{tab:osc}}. At the far detector, muon
disappearance probability depends on standard oscillation parameters $\theta
_{23}$ and $\Delta m_{31}^{2}$ and NSI parameters $\varepsilon _{\mu \mu }$, 
$\varepsilon _{\mu \tau }$, $\varepsilon _{\tau \mu }$ and $\varepsilon
_{e\mu }$. Electron appearance probability at the far detector depends on
all the standard oscillation parameters $\theta _{23}$, $\theta _{12}$, $%
\theta _{13}$, $\Delta m_{31}^{2}$, $\Delta m_{21}^{2}$ and $\delta _{CP}$
and also on NSI parameters $\varepsilon _{\mu e}$ and $\varepsilon _{\tau e}$%
. Dependence of the appearance and disappearance probabilities to the
standard oscillation and NSI parameters are listed in table \ref{tab:osc}.
For the complete approximate analytical expression of all the oscillation
probabilities formulas see Ref. \cite{Kopp:2007ne}.

As a first exercise to see the effects of CC\ NSIs on the oscillation
probabilities at the far detector of the DUNE with baseline of 1300 km are
shown in Fig. 1. In Fig 1(a) the electron neutrino appearance probability is
shown for the standard interaction and NSIs when only one of the NSI
parameters is nonzero and is equal to 0.1. In Fig. 1(b) the disappearance of
the muon neutrino is shown. Notice that the values for the standard
oscillation parameters and their uncertainties were taken from nu-fit \cite%
{GonzalezGarcia:2012sz,Gonzalez-Garcia:2014bfa} with $\delta _{CP}=0$ and
the resolution of the plot along the energy axis was taken as 0.1 MeV. 
\begin{figure}[tbp]
\begin{center}
\subfigure[]{\includegraphics[width=0.49\textwidth]{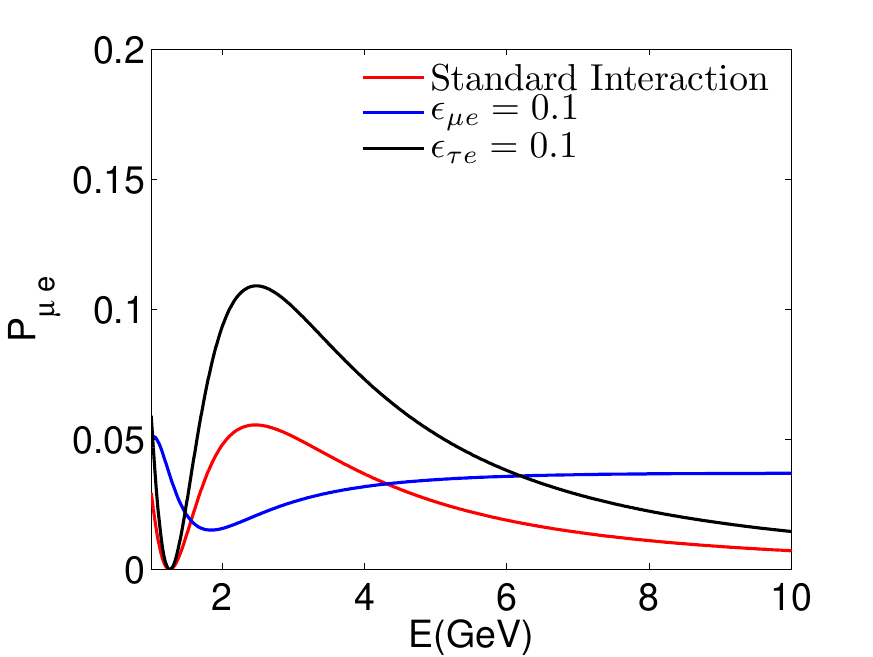}} %
\subfigure[]{\includegraphics[width=0.49\textwidth]{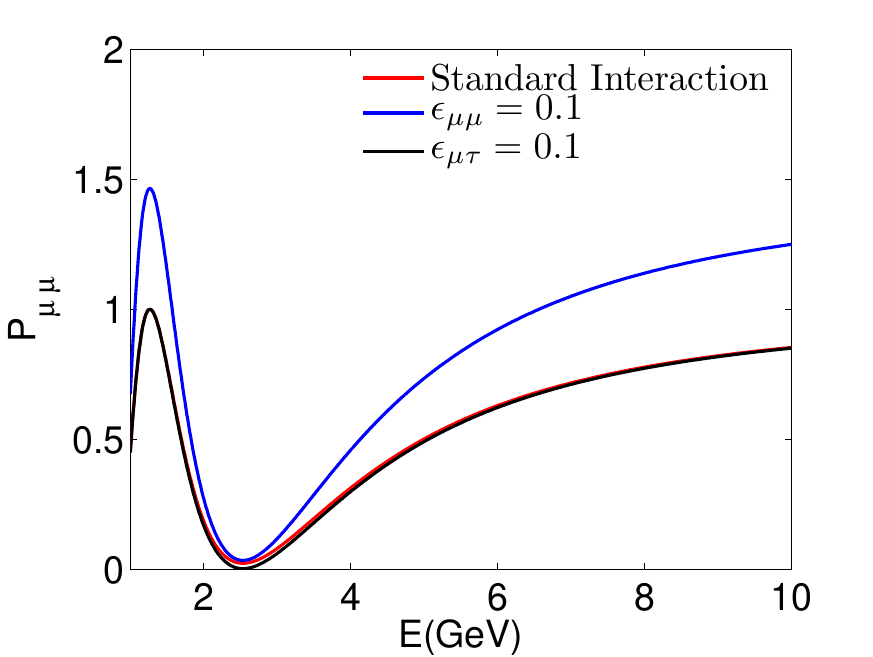}}
\end{center}
\par
\vspace{0.0cm}
\caption{An illustration of oscillation probabilities at far detector of
baseline 1300km for muon neutrino disappearance and electron neutrino
appearance. The standard oscillation parameters are taken from nu-fit 
\protect \cite{GonzalezGarcia:2012sz,Gonzalez-Garcia:2014bfa} and $\protect%
\delta _{CP}=0$ . For NSIs curve, only value of one NSI parameters
considered to be nonzero and equal to 0.1. The value of all the phases are
set equal to zero.}
\end{figure}

\section{Characteristics of DUNE and the Analysis details\label{anal}}

The DUNE will be a long baseline accelerator based neutrino oscillation
experiment where neutrinos/antineutrino beams will be produced at FermiLab
and will be detected at Sanford Underground Research Facility, 1300 km away
from the FermiLab. In our simulations, we consider 1.2 MW proton beam which
produces the neutrino and antineutrino beam from pion decays with the
neutrino energy ranging from 100 MeV to 20 GeV and with the peak around 3
GeV. We take the spectrum from Ref. \cite{spectrum} and consider the
reference beams for the near and far detector in neutrino and antineutrino
modes with 3 years of data taking in each mode. We consider a Liquid Argon
Time-Projection Chamber (LArTPC) with 34 kton fiducial mass for the far
detector. The details of the near detector are under discussions, however,
we consider a near detector with 5 ton fiducial mass placed at 460 m
baseline. The NC events, lepton flavor misidentication and the intrinsic
background are considered as the main sources of background, while
neglecting the other background sources.

\begin{figure}[tbp]
\begin{center}
\subfigure[]{\includegraphics[width=0.49\textwidth]{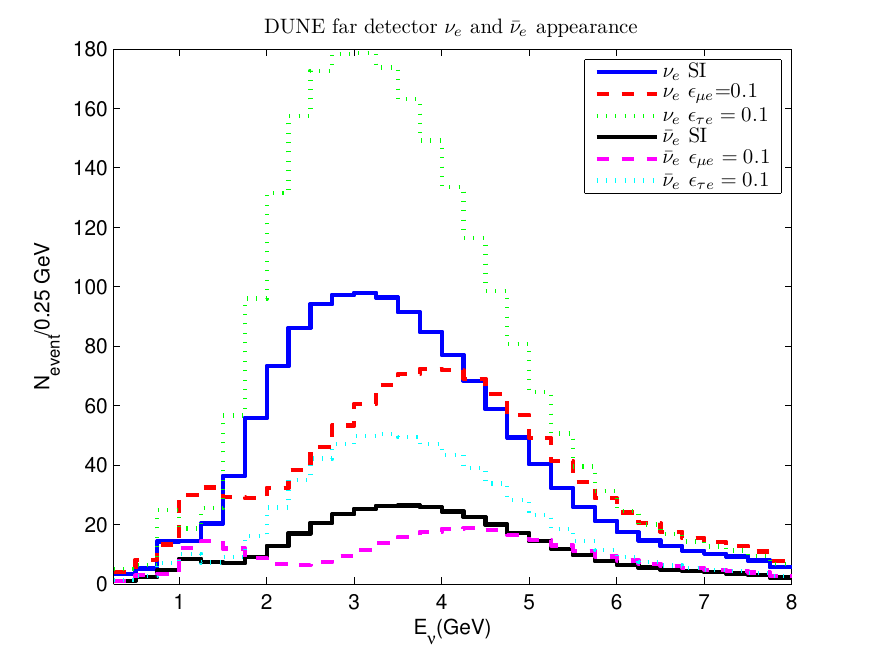}} %
\subfigure[]{\includegraphics[width=0.49\textwidth]{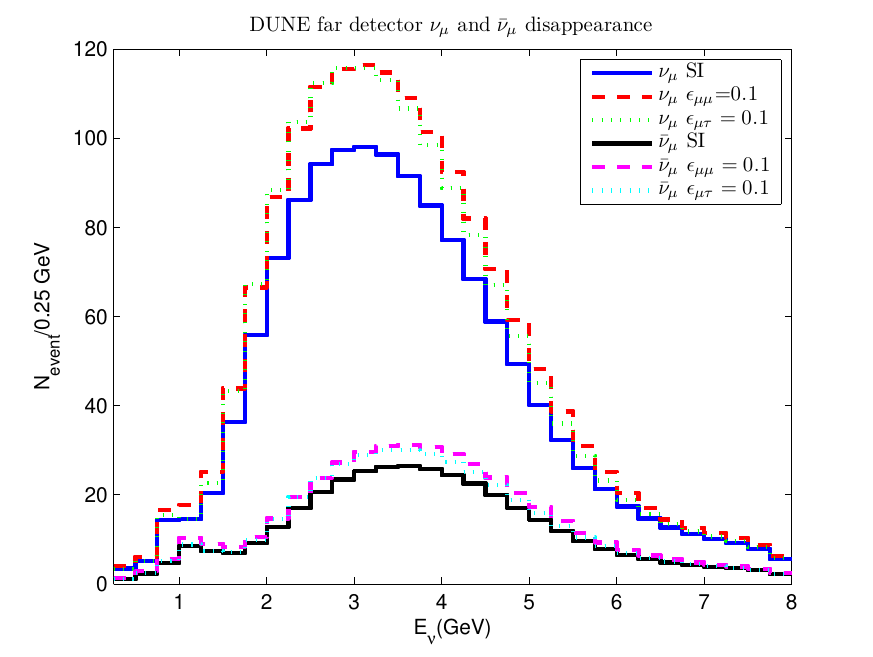}} %
\subfigure[]{\includegraphics[width=0.49\textwidth]{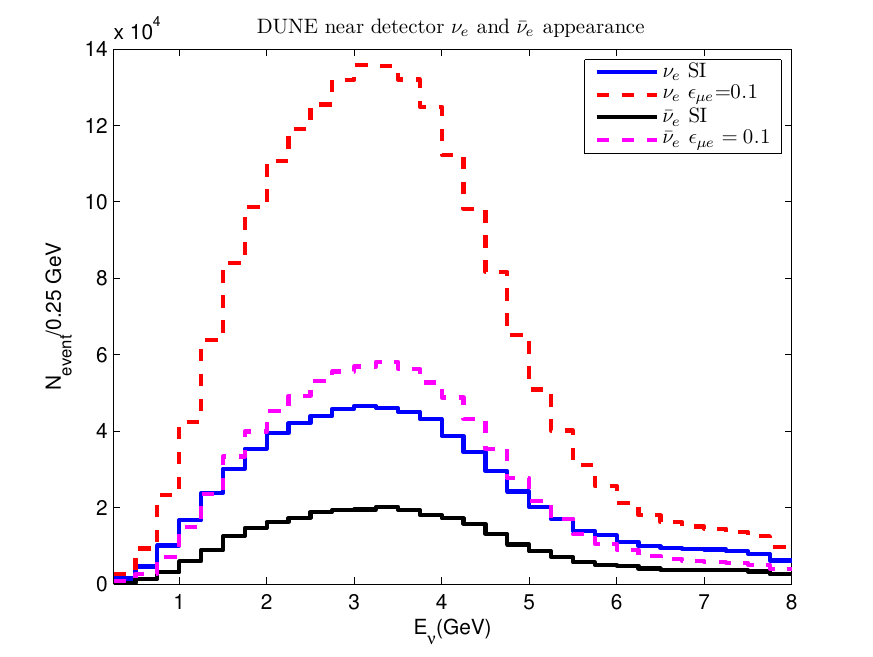}} %
\subfigure[]{\includegraphics[width=0.49\textwidth]{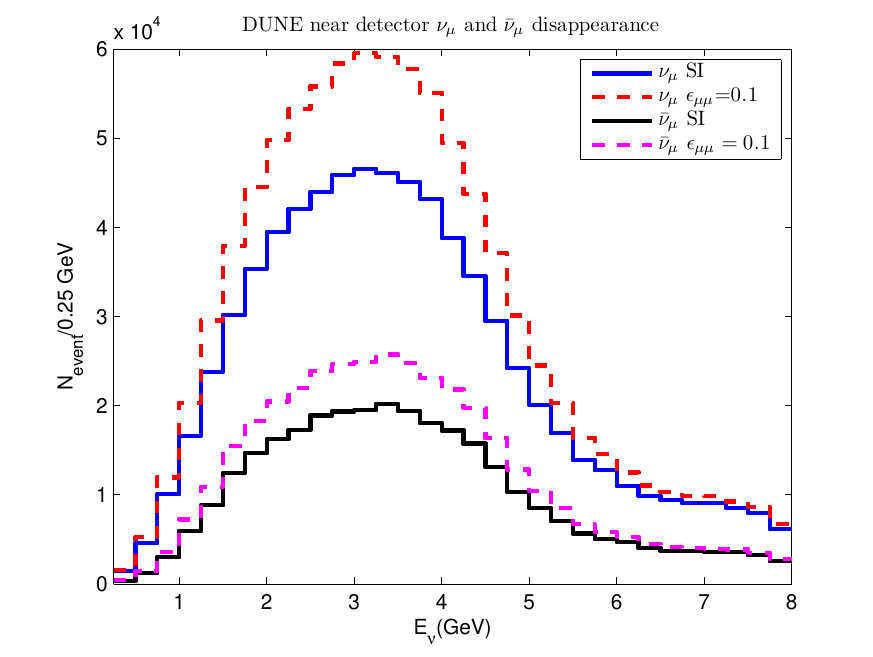}}
\end{center}
\par
\vspace{0.0cm}
\caption{Number of events per bin for near and far detector of DUNE for
three years of data taking in each mode. Number of events includes both
signal and background. Oscillation parameters are taken from nu-fit 
\protect \cite{GonzalezGarcia:2012sz,Gonzalez-Garcia:2014bfa}. $\protect%
\delta _{CP}$ is considered equal to zero. Detection efficiencies are taken
from Ref. \protect \cite{Adams:2013qkq} and neutrino flux from Ref. 
\protect \cite{spectrum}. For NSIs curve, only value of one NSI parameters
considered to be nonzero and equal to 0.1. The value of all the phases are
set equal to zero.}
\label{NOE}
\end{figure}

We consider the energy resolutions for the CC\ detections as $15\%/\sqrt{%
E(GeV)}$ for $\nu _{e}\ $and $20\%/\sqrt{E(GeV)}$ for $\nu _{\mu },$ while
the efficiency of CC detection of $\nu _{e}$ and $\nu _{\mu }$ as 80$\%$ and
85$\%$ , respectively. Both the NC and Lepton flavor misidentification rates
are taken 1$\%$. The flux uncertainty is taken 5$\%$, the calibration error
is 2$\%$ and the flux uncertainty of the background is 10$\%$. We take the
detector performance from Table 4.2 of Ref. \cite{Adams:2013qkq}. For the
analysis, we consider the energy range between 0.25 GeV to 8 GeV and 31 bins
in the unit steps of 0.25 GeV. The CC and NC cross sections are taken from
Refs. \cite{Messier:1999kj, Paschos:2001np}. The number of events in each
bin for the standard three neutrino oscillation and in the case of NSI for\ $%
\varepsilon _{\mu e}$= 0.1 and $\varepsilon _{\mu e}$= 0.1 in the appearance
channel and with $\varepsilon _{\mu \mu }$= 0.1 and $\varepsilon _{\tau e}$=
0.1 in the disappearance channel for the near and far detector data are
shown in Fig. \ref{NOE}, where the contribution in each bin is the aggregate
of both signal and background events. In the near detector appearance mode,
almost all of the events are background events since the number of events of
the signal is negligible. The values with their uncertainties of the
oscillation parameters are taken from nu-fit \cite%
{GonzalezGarcia:2012sz,Gonzalez-Garcia:2014bfa} and the value of the CP
phase is taken zero. For matter density profile, we use PREM with 5\%
uncertainties \cite{PREM}. Our number of events for near and far detector
are in agreement with the result of Refs. \cite%
{deGouvea:2015ndi,Choubey:2016fpi}, except for the the slight differences
which occur due to the differences in the neutrino fluxes, details of the
detector performance, oscillation parameters such as the different value of $%
\delta _{CP}$. For the result of the near detector, in comparison with Ref. 
\cite{Choubey:2016fpi}, we consider baseline of the near detector equal to
460 m, while they consider 595 m. Moreover, our number of events include
both signal and background while they consider only the signal. For
simulation and statistical inference, we used GloBES software \cite%
{Huber:2004ka,Huber:2007ji}. The chi-squared test is used by GloBES for
statistical inferences and a Gaussian error is implied by the software. For
considering the CC NSIs, we include the code from Refs. \cite%
{Kopp:2007ne,Kopp:2006wp}. We consider the uncertainties of source and
detector CC NSIs from Ref. \cite{Biggio:2009nt}.

\section{Results and Discussion\label{result}}

In this section we discuss the main results of the effects of CC NSIs on the
determination of the standard parameters, $\delta _{CP}$ and $\theta _{23}$,
fake CP-violation due to the CC NSIs and the constraints on the source and
detector CC NSI parameters at the near and far detectors of DUNE. Since the
hierarchy measurements are mainly affected by the NC NSIs \cite%
{Bakhti:2014pva, Coloma:2016gei} in the long baseline experiments like DUNE
and the CC NSIs play no important role in the determination of the correct
mass ordering, so in our case assuming either mass ordering will not affect
our results. For our analysis we consider the normal mass ordering as the
true hierarchy.

\subsection{Effects of CC\ NSIs on simultaneous determination of $\protect%
\delta _{CP}$ and octant of $\protect \theta _{23}$ at DUNE}

To investigate the effects of source and detector CC NSIs on the
determination of $\delta _{CP}$ and octant of $\theta _{23}$, we simulate
our data by considering two choices of the value of $\delta _{CP}$. First,
we take $\delta _{CP}=0,$ which corresponds to the CP-conserving case and
second, we consider $\delta _{CP}=270^{\circ }$, which is the maximal
CP-violation in the standard neutrino oscillation scheme and is determined
as the global best fit value \cite{Gonzalez-Garcia:2014bfa}. With these two
extreme values of $\delta _{CP}$, we find the correlation from the two
parameter fits of $\delta _{CP}$ vs. $\theta _{23}$ in the standard
oscillation scheme where we set all of the other NSI parameters equal to
zero and minimize over all the other standard oscillation parameters in
these fits. In the second case, we repeat the two parameters fit analysis
and minimize over both the standard mixing parameters and also over all of
the source and detector NSI parameters, $|\varepsilon _{\alpha \beta }|$ and 
$\phi _{\alpha \beta }$. For minimization over the NSI parameters, we
consider the current uncertainty of these parameters from Ref. \cite%
{Biggio:2009nt,ANK3}.

\begin{figure}[tbp]
\begin{center}
\subfigure[]{\includegraphics[width=0.49\textwidth]{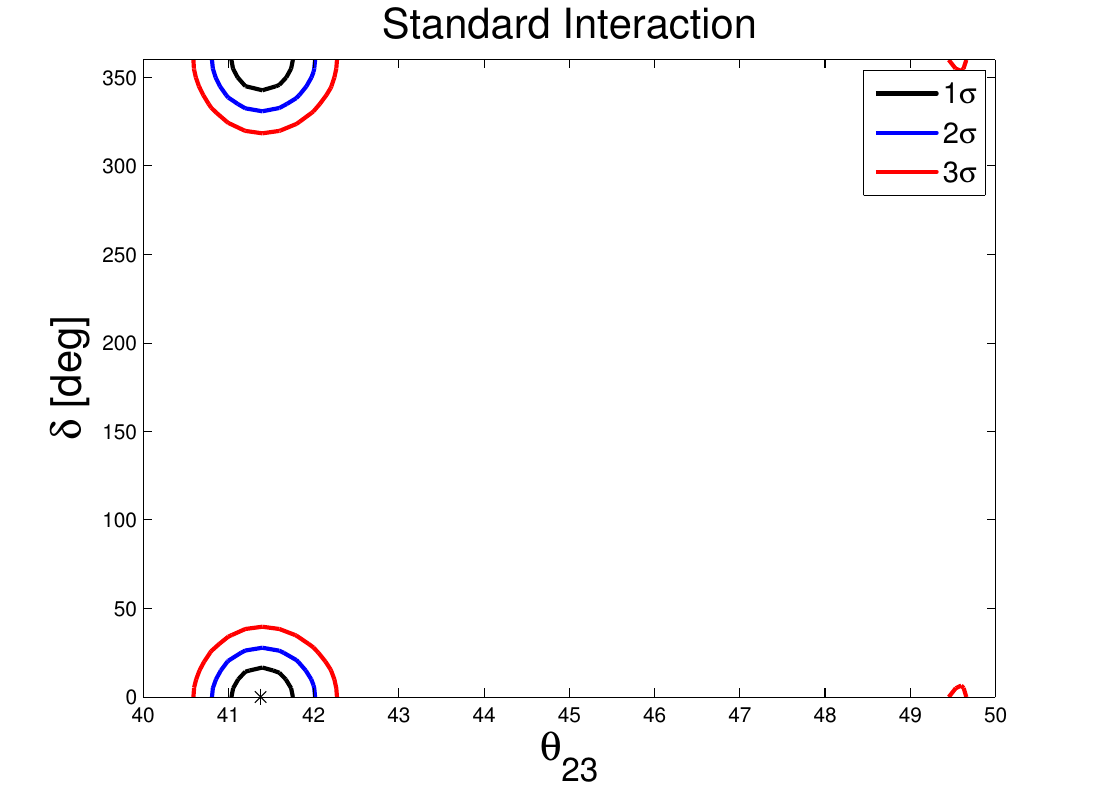}} %
\subfigure[]{\includegraphics[width=0.49\textwidth]{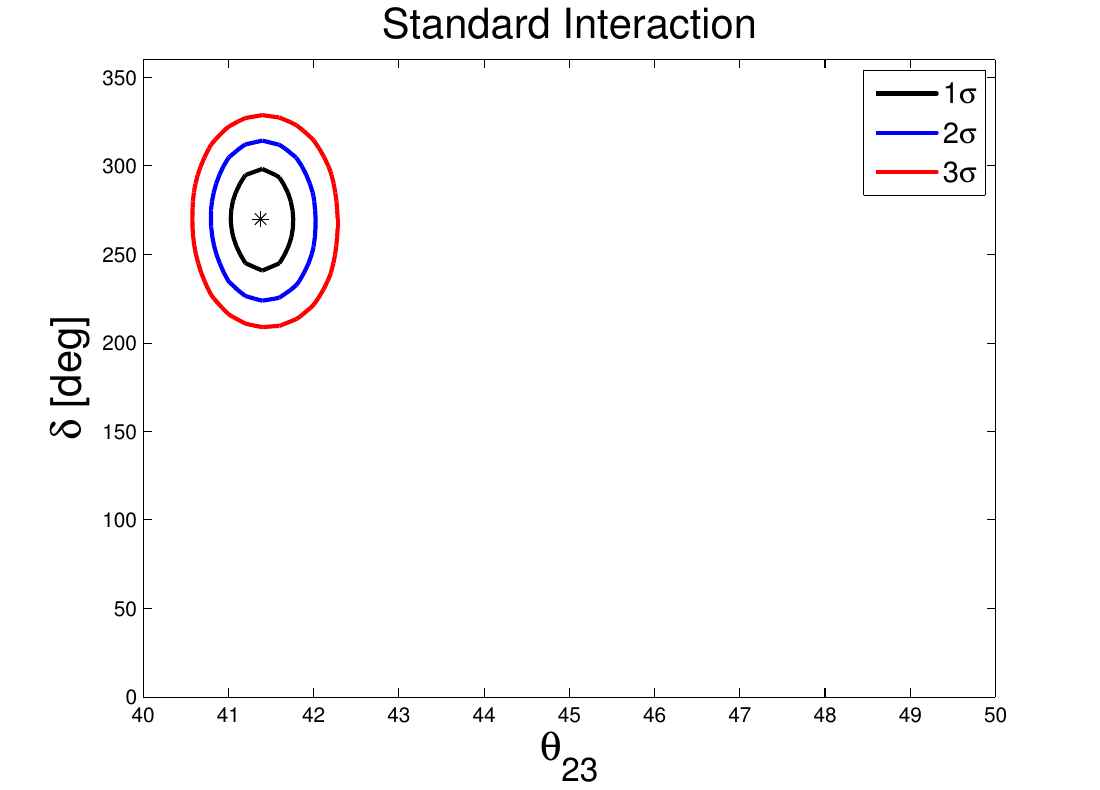}} %
\subfigure[]{\includegraphics[width=0.49\textwidth]{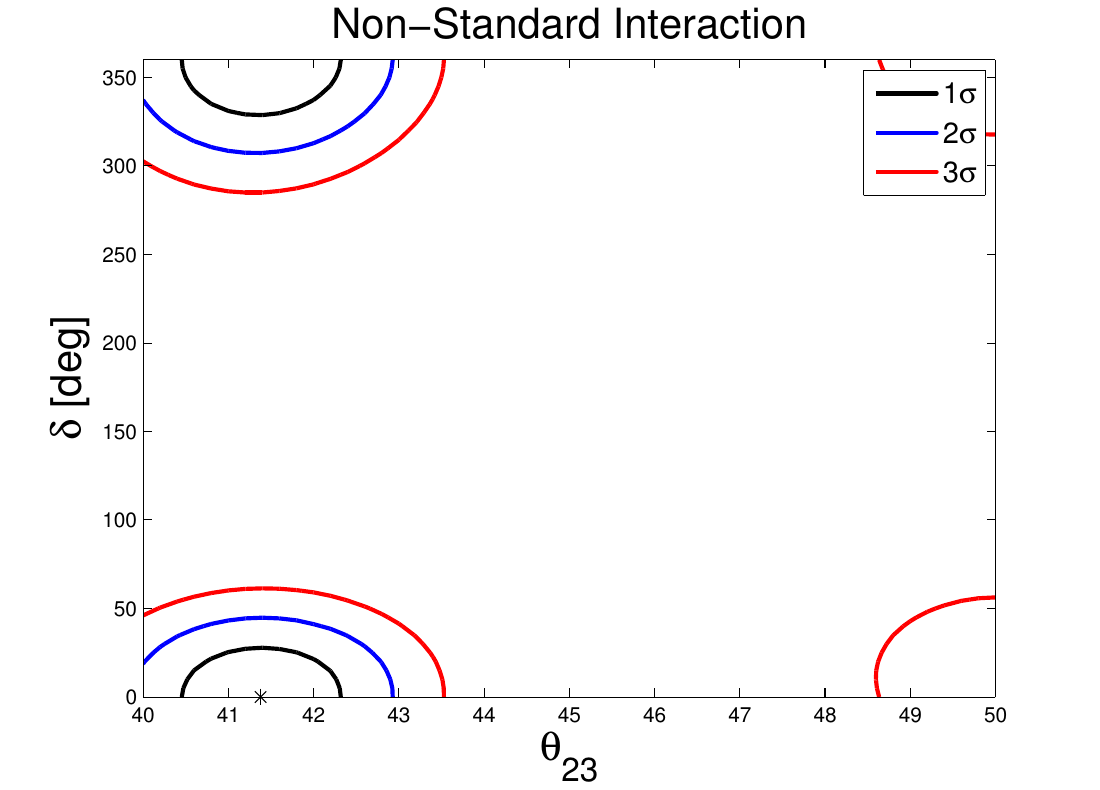}} %
\subfigure[]{\includegraphics[width=0.49\textwidth]{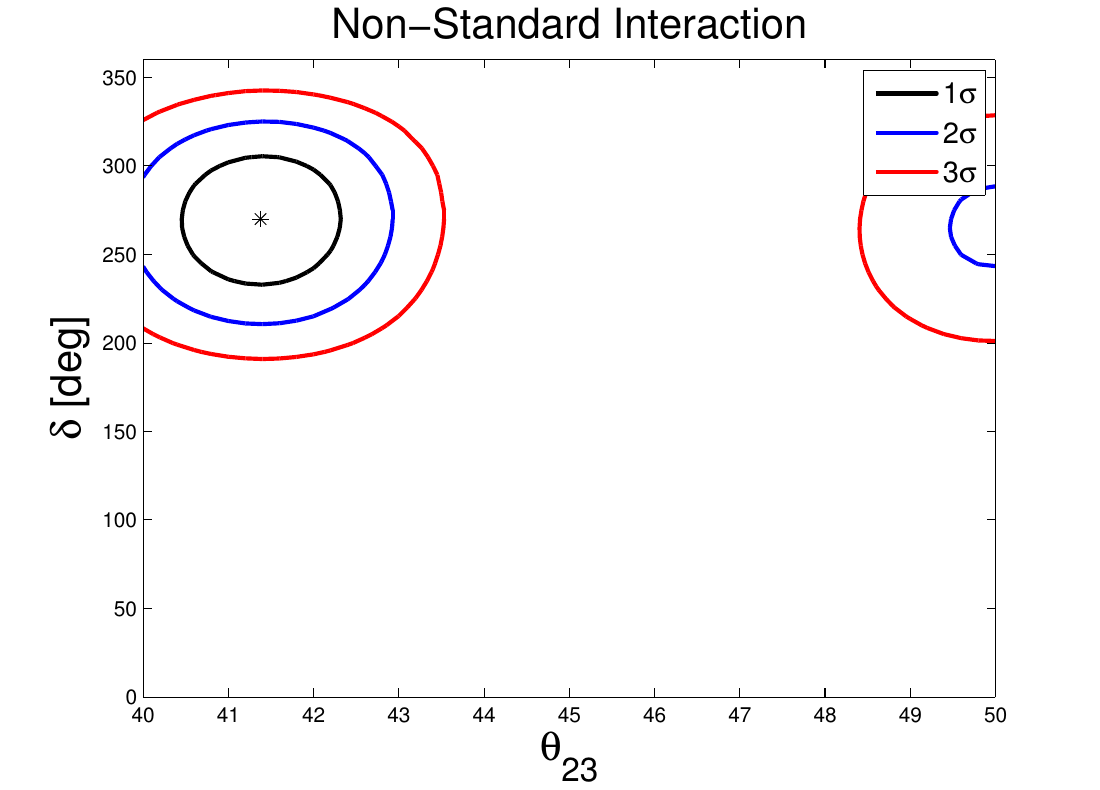}}
\end{center}
\par
\vspace{0cm}
\caption{Simultaneous determination of $\protect \delta _{CP}$ and octant of $%
\protect \theta _{23}$ by DUNE after six years of data taking, three years in
each mode, for $\protect \delta _{CP}=0^{\circ }$ for panels a and c and $%
\protect \delta _{CP}=270^{\circ }$ for panels b and d. In panels a and b, we
marginalized over the standard oscillation parameters, while in panels c and
d, we marginalized over standard and NSI source and detector parameters.}
\label{T23CP}
\end{figure}

Using the characteristics of DUNE and the analysis details as explained in
section III, we show the results for the analysis of the simultaneous
measurements of $\theta _{23}$ and $\delta _{CP}$ in Fig. \ref{T23CP}. Fig. %
\ref{T23CP}(a) and Fig. \ref{T23CP}(b) have been obtained when we minimize
only over the standard oscillation parameters while set all NSI parameters
equal to zero. On the other hand, Fig. \ref{T23CP}-c and Fig. \ref{T23CP}-d
have been obtained when we minimized over both the standard oscillation
parameters and over all of the source and detector NSI parameters. Fig. \ref%
{T23CP}(a) and Fig. \ref{T23CP}-c correspond to the simulated data with $%
\delta _{CP}=0^{\circ }$, while Fig. \ref{T23CP}-b and Fig. \ref{T23CP}-d
correspond to the $\delta _{CP}=270^{\circ }$ case.

By comparing Fig. \ref{T23CP}-a with Fig. \ref{T23CP}(c) for the $\delta
_{CP}=0^{\circ }$ and Fig. \ref{T23CP}-b with Fig. \ref{T23CP}(d) for the $%
\delta _{CP}=270^{\circ }$, the results show that the 3$\sigma $ C.L.
determination of $\theta _{23}$ (Fig. \ref{T23CP}(a) and Fig. \ref{T23CP}%
(b)) in the standard scenario is destroyed in the presence of CC NSIs (Fig.
3(c) and Fig. 3(d)) within the currently known constraints of the CC NSI
parameters from Ref. \cite{Biggio:2009nt,ANK3}. When the uncertainties of
the source and detector CC NSIs are ignored, the octant of $\theta _{23}$
can be determined with 3$\sigma $ C.L. as shown in Fig. \ref{T23CP}(a) and
Fig. \ref{T23CP}(b), while when these uncertainties are taken into account,
the octant cannot be determined with 3$\sigma $ C.L., and the 2$\sigma $ and
3$\sigma $ significance levels get affected due to the CC\ NSI
contributions. Similar deterioration of the 3$\sigma $ C.L. determination of 
$\delta _{CP}$ within the standard scenario takes place for both types of
data when the CC NSIs are included as can be seen by comparing Fig. \ref%
{T23CP}(a) with Fig. \ref{T23CP}(c) and Fig. \ref{T23CP}(b) with Fig. \ref%
{T23CP}(d). The accuracy of the $\delta _{CP}$ measurement gets worsen
approximately 80$\%$ for $\delta _{CP}=0^{\circ }$ and 50$\%$ for $\delta
_{CP}=270^{\circ }$.

\subsection{Effects of the CC\ NSI fake CP-violation parameter $(\protect%
\delta _{CP}^{^{\prime }})$}

\begin{figure}[tbp]
\begin{center}
\subfigure[]{\includegraphics[width=0.32\textwidth]{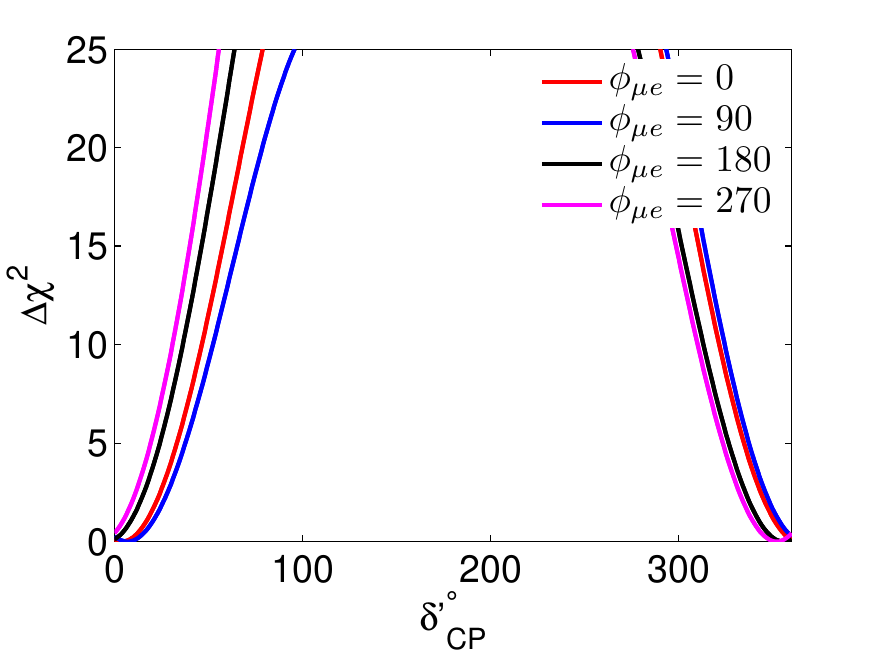}} %
\subfigure[]{\includegraphics[width=0.32\textwidth]{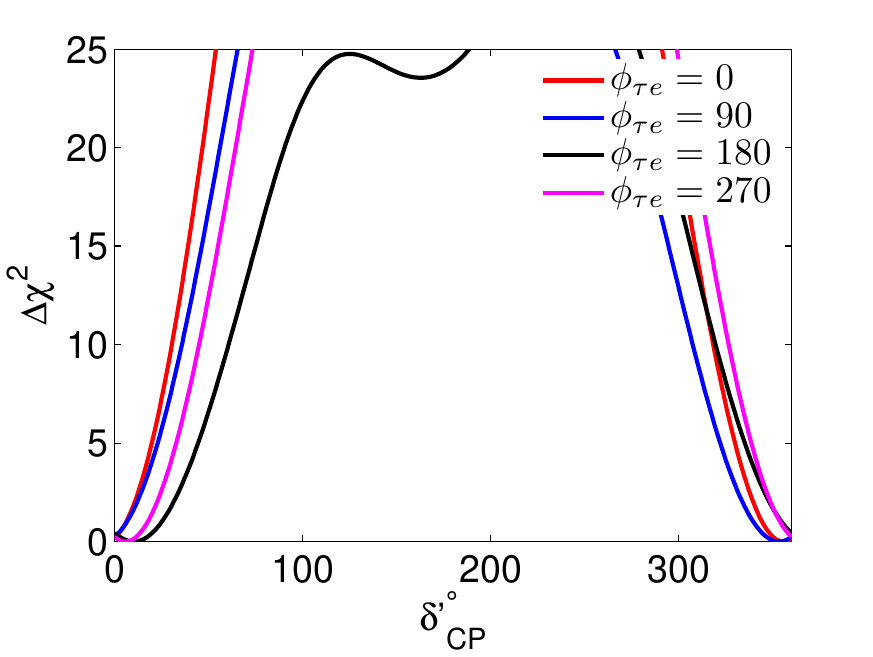}} %
\subfigure[]{\includegraphics[width=0.32\textwidth]{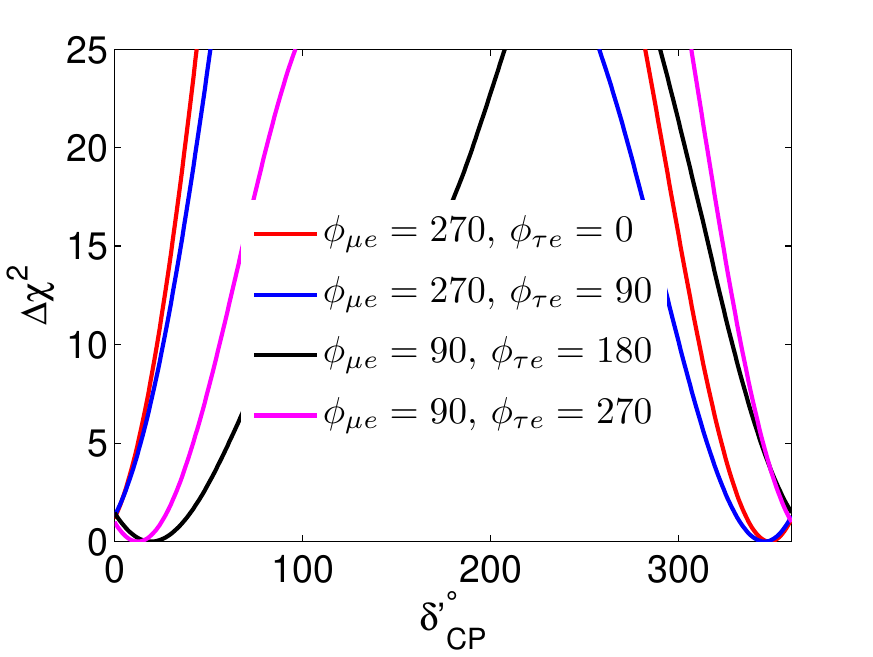}}
\end{center}
\par
\vspace{0cm}
\caption{Induced fake CP-violation phase $\protect \delta _{CP}^{\prime }$ in
DUNE by considering nonzero source and detector NSIs. In panels a and b only 
$|\protect \epsilon _{\protect \mu e}|$ and $|\protect \epsilon _{\protect \tau %
e}|$ is nonzero and is set equal to 0.025 respectively and the values of
corresponding NSI phases are 0$^{\circ }$, 90$^{\circ }$, 180$^{\circ }$ and
270$^{\circ }$. All the other NSI parameters are set equal to zero. In panel
c all of these NSI parameters are considered nonzero and equal to 0.025. The
values of corresponding NSI phases are shown in the legends.}
\label{DeltaCP}
\end{figure}

Due to the importance of $\delta _{CP}$ and its explicit measurement program
at DUNE we evaluate numerically the impacts of source and detector CC NSI
phases ($\phi _{\alpha \beta }$) which can mimic the $\delta _{CP}$
measurement resulting into the fake CP-violation ($\delta _{CP}^{^{\prime }}$%
)\textit{, }where $\delta _{CP}^{^{\prime }}$\ is the measured value of the
CP-phase by DUNE which includes the contribution from the standard and
nonstandard CP-violating phases. When $\delta _{CP}\ $= 0, any nonzero
measured value of\ $\delta _{CP}^{^{\prime }}$ could be induced by the
nonzero value of NSI parameters, $\varepsilon _{\alpha \beta }$ and $\phi
_{\alpha \beta }$. In Fig. \ref{DeltaCP}, we show the one parameter fit of $%
\delta _{CP}^{^{\prime }}$ results for the various choices of the nonzero
values of CC NSI moduli and phases for the special case of $\delta
_{CP}=0^{\circ }$. We first take the value of $|\varepsilon _{\mu e}|$ and $%
|\varepsilon _{\tau e}|$ individually as nonzero and then take all the three
nonzero at a time, while set them equal to 0.025 in all the cases \cite%
{Biggio:2009nt, ANK3}. In this analysis we consider four values 0$^{\circ }$%
, 90$^{\circ }$, 180$^{\circ }$ and 270$^{\circ }$ for the corresponding
phases in each case. Fig. \ref{DeltaCP}(a) and Fig. \ref{DeltaCP}(b) have
been obtained when $|\varepsilon _{\mu e}|=\ $0.025 and $|\varepsilon _{\tau
e}|=\ $0.025, respectively at a time while set the other two parameters
equal to zero. The four different choices of the three CC\ NSI phases are
shown through different color legends in the figure. In Fig. \ref{DeltaCP}%
(c), both absolute parameters $|\varepsilon _{\mu e}|$ and $|\varepsilon
_{\tau e}|$ are taken nonzero and set equal to 0.025. Notice that we have
minimized over all the standard mixing parameters in this analysis.

The results show that in some cases of the parameter choices, for instance,
for black curve of Fig. \ref{DeltaCP}(c), we obtained the best-fit value of $%
\delta _{CP}^{\prime }$as 30$^{0}$ more than $80\%$ C.L.. Similarly, in
general, we can see from all the three panels of Fig. \ref{DeltaCP} that the
absolute value of $\delta _{CP}^{^{\prime }}$ varies from $-50^{\circ }$ to $%
50^{\circ }$. The analysis shows that source and detector CC NSIs parameters
can induce a significant amount of fake CP-violations with in the current
best limits obtained in Ref. \cite{Biggio:2009nt,ANK3}. 

\subsection{Sensitivities to source and detector CC\ NSIs at DUNE}

We further analyze the simulated data for the DUNE setup to constrain the CC
NSI parameters at the source and detector. For this, we simulate our data
with all the best fit values of the standard mixing parameters from nu-fit 
\cite{Gonzalez-Garcia:2014bfa} and find the single parameter fits for the
relevant absolutes of the NSI parameters while set all the other NSIs
parameters to zero. The results of this study are demonstrated in Fig. \ref%
{Fig:epsilon} and the bounds extracted at 90$\%$ C.L. are given in table \ref%
{Tab:epsilon}. We take 90$\%$ C.L. projection over the distributions of the
single parameter fits to extract the bounds as shown in Fig. 5. We use all
the neutrino and antineutrino oscillation channels relevant for DUNE at both
the near and far detectors. In the fourth column of table \ref{Tab:epsilon},
we also show the current bounds from the Ref. \cite{Biggio:2009nt,nuno} for
comparison. By the general comparison of bounds from this work with the
bounds from Ref. \cite{Biggio:2009nt,nuno} (fourth column of table II) shows
that for the 3+3 years of running, DUNE has a potential to give more
stringent bounds than the existing ones from the other experiments \cite%
{ANK2,ANK3}. Interestingly, the bounds for the relevant parameters at the
near detector has an overall one order of magnitude improvement than all of
the current bounds, while using the data of the far detector, there is an
overall a factor of 2 improvement except for $\varepsilon _{\mu \tau }$, $%
\varepsilon _{\tau e}$ and $\varepsilon _{\tau \mu }$ parameters which are
weakly constrained in comparison to the current bounds. 

As discussed in section \ref{theo}, the near detector is sensitive to $%
\varepsilon _{ee}$, $\varepsilon _{\mu \mu }\ $and$\  \varepsilon _{\mu e}$,
but not to $\varepsilon _{\mu \tau }$, $\varepsilon _{e\mu }$, $\varepsilon
_{\tau \mu }\ $and $\varepsilon _{\tau e}$, thus the former set of parameter
can be constrained at the near detector as shown in Fig.5(a, b, c), while
the latter cannot be constrained as can be seen in Fig.5(d, e, f, g). As can
be seen from TABLE II, constraints on $\varepsilon _{\mu \tau }$, $%
\varepsilon _{e\mu }$, $\varepsilon _{\tau e}$, $\varepsilon _{\tau \mu }$
at the far detector are stronger in some cases while weaker in the others.
For the far detector, when we fix $\delta _{CP}$ or $\theta _{23}$ to their
central values and try to find correlation between the NSI parameters $%
\varepsilon _{\mu \tau }$, $\varepsilon _{e\mu }$, $\varepsilon _{\tau e}$, $%
\varepsilon _{\tau \mu }$ and $\delta _{CP}$ or $\theta _{23}$, we find that
there is a strong correlation between $\varepsilon _{\tau e}$ and $\delta
_{CP}$ and the constraint on this parameter at 90$\%$ C.L. is 0.042. The
same correlation, although weaker, also exists for $\varepsilon _{e\mu }$.
Similar correlations also exist between $\varepsilon _{\mu \tau }$, $%
\varepsilon _{\tau \mu }$ and $\theta _{23}$, where the bounds at the 90$\%$
C.L. are 0.013 and 0.014, respectively. This indicates that the weaker
bounds of some of the parameters are due to the existing uncertainties in $%
\delta _{CP}$ and $\theta _{23}$. 
\begin{figure}[tbp]
\begin{center}
\subfigure[]{\includegraphics[width=0.32\textwidth]{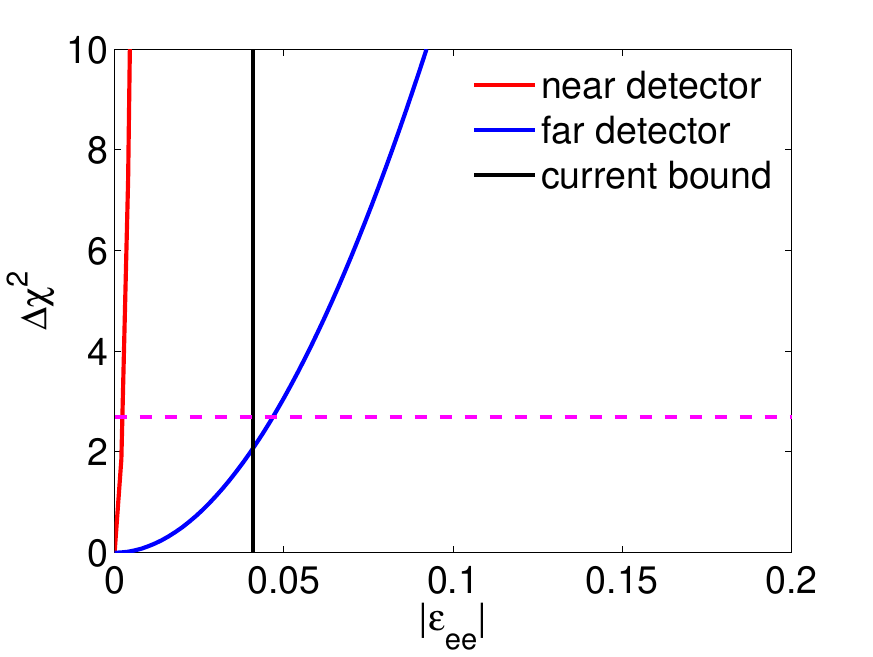}}%
\subfigure[]{\includegraphics[width=0.32\textwidth]{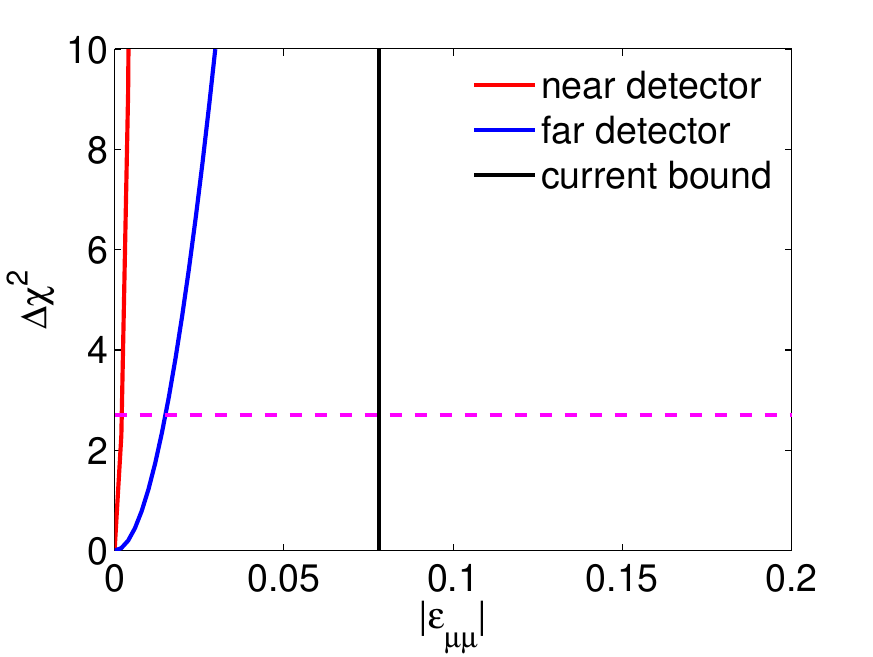}}%
\subfigure[]{\includegraphics[width=0.32\textwidth]{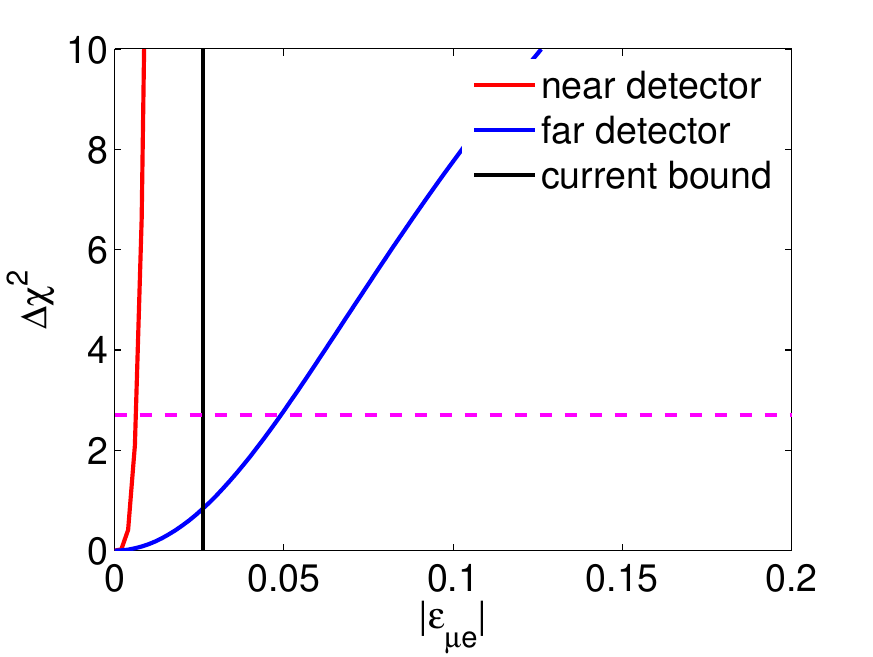}} \  \  \  \  \ %
\subfigure[]{\includegraphics[width=0.32\textwidth]{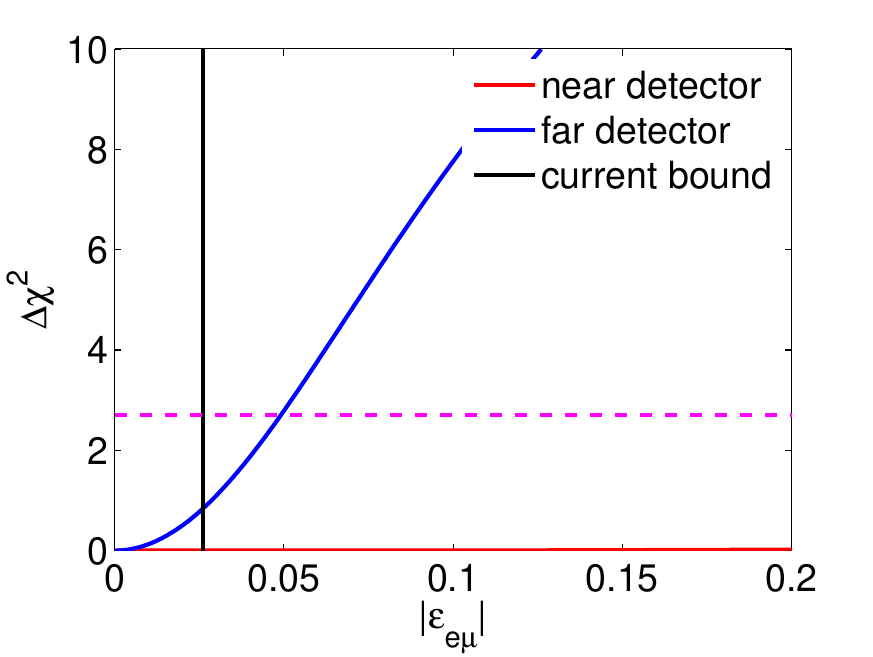}}%
\subfigure[]{\includegraphics[width=0.32\textwidth]{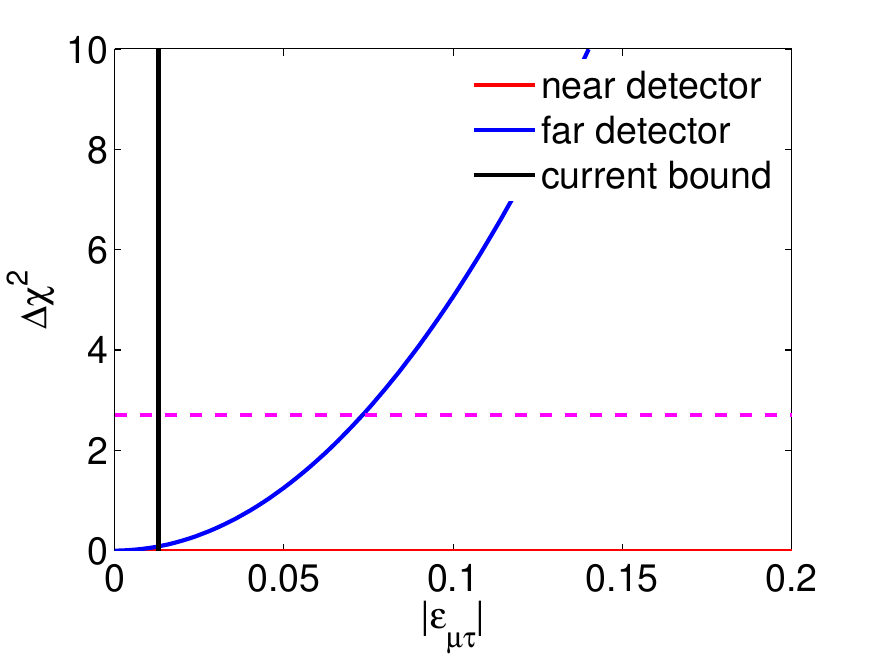}} \ %
\subfigure[]{\includegraphics[width=0.32\textwidth]{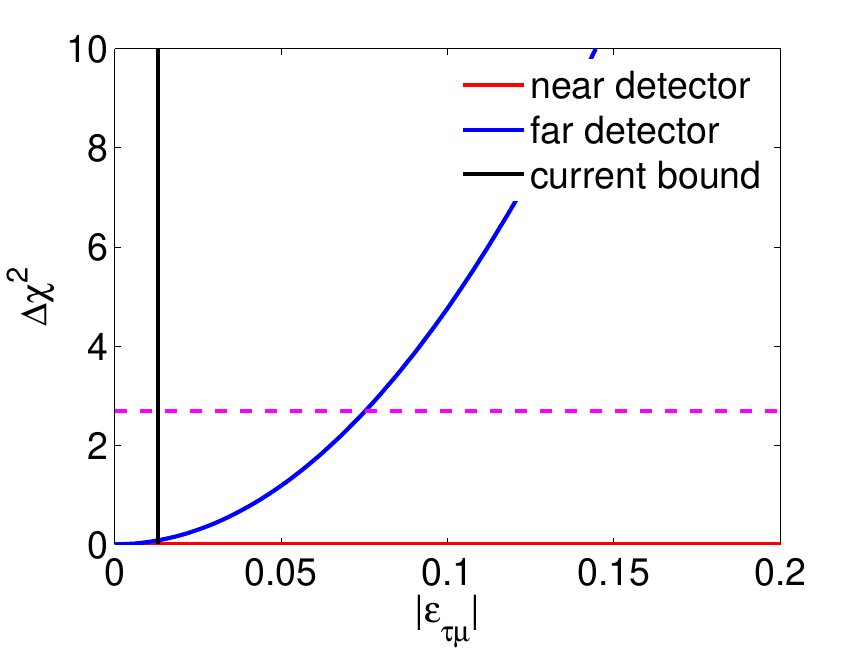}}%
\subfigure[]{\includegraphics[width=0.32\textwidth]{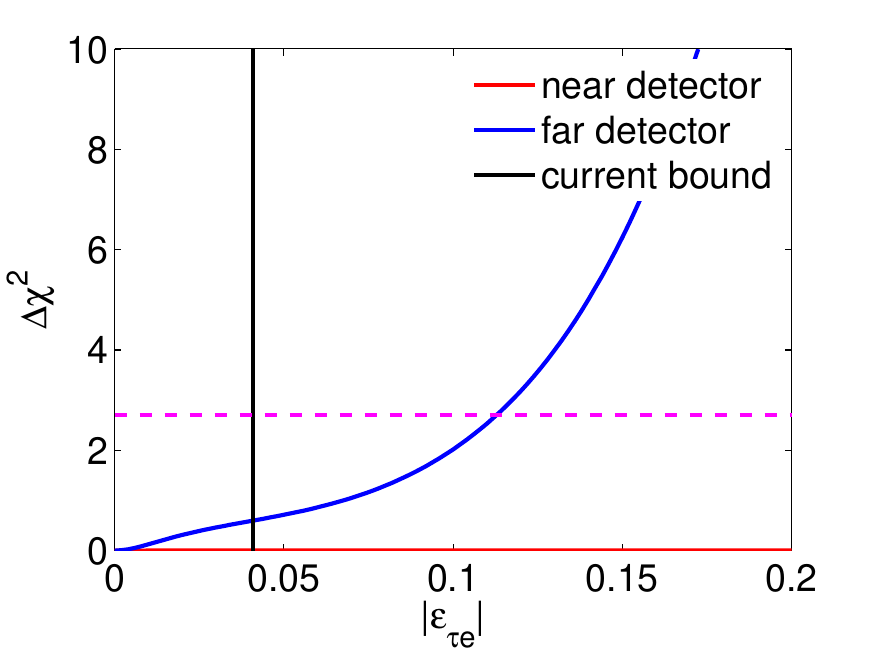}}
\end{center}
\caption{Constraints on source and detector NSI parameters by near and far
detector of DUNE. The current "indirect" bounds at the 90$\%$ C.L. from Ref. 
\protect \cite{Biggio:2009nt} for each parameter are shown by the black lines
for comparison. The 90$\%$ C.L. projections are shown by dashed lines.}
\label{Fig:epsilon}
\end{figure}

As a check we also repeat the exercise of the analysis of Fig. \ref{T23CP}
using the bounds obtained from this study for the near detector case and
found that there are marginal differences from the results of Fig. \ref%
{T23CP}. The reason is that the near detector cannot constrain the $%
\varepsilon _{\tau e}$ parameter and, as shown in the subsection A above,
this parameter is central to the appearance channels and has a large impact
on the measurement of $\delta _{CP}$ and octant of $\theta _{23}$. Notice
that the near detector will constrain $\varepsilon _{\mu e}$. If these
parameters have values $O(0.01)$, their effects will be detectable in the
near detector. Since the near detector is sensitive to $\varepsilon _{\mu e}$%
, the near detector can be more helpful to discriminate between $\delta
_{CP}^{\prime }$ and $\delta _{CP}$ than the far detector. Also near
detector has no sensitivity to $\varepsilon _{\tau e}$, so it cannot
discriminate between $\delta _{CP}^{\prime }$ and $\delta _{CP}$ in all the
cases. We do similar simulations for Fig. 4 with the near detector bounds
obtained in this work, and in case similar to Fig.4(a) the induced fake CP
phase is negligible, $\delta _{CP}^{\prime }=0$. In the case similar to
Fig.4(b), since the near detector do not constrain $\varepsilon _{\tau e}$,
the result is the same as Fig.4(b), and in the case of Fig.4(c), the effect
of induced fake CP phase is less than half of the result demonstrated in
Fig.4(c).

\begin{table}[tbp]
\begin{center}
\begin{tabular}{|c|c|c|c|}
\hline
Parameters & Far Detector & Near Detector & Current Constraints \\%
[0.5ex] \hline \hline
$|\epsilon _{ee}|$ & 0.046 & 0.003 & 0.041 \\ \hline
$|\epsilon _{\mu \mu }|$ & 0.015 & 0.002 & 0.078 \\ \hline
$|\epsilon _{\mu e}|$ & 0.009 & 0.006 & 0.026 \\ \hline
$|\epsilon _{\mu \tau }|$ & 0.074 & - & 0.013 \\ \hline
$|\epsilon _{e\mu }|$ & 0.049 & - & 0.026 \\ \hline
$|\epsilon _{\tau \mu }|$ & 0.076 & - & 0.013 \\[0.5ex] \hline
$|\epsilon _{\tau e}|$ & 0.113 & - & 0.041 \\ \hline
\end{tabular}%
\end{center}
\caption{One degree of freedom constraints on the CC NSI parameters at the
near and far detector of DUNE, obtained from Fig. \protect \ref{Fig:epsilon}
at 90$\%$C.L., while the current "indirect" bounds from Ref. \protect \cite%
{Biggio:2009nt,nuno} at 90$\%$ C.L. also given in fourth row for comparison.
The blank entries indicate that the DUNE near detector has no sensitivity to
these parameters.}
\label{Tab:epsilon}
\end{table}

\section{Summary and Conclusions\label{sum}}

In this paper, we have explored the effects of CC NSIs at DUNE. We have
studied the impacts on the simultaneous measurements of the $\delta _{CP}$
and $\theta _{23}$ due to the CC NSIs at neutrino source and at the
detector. We study how the statistical significance level of the measurement
of $\delta _{CP}$ and $\theta _{23}$ at the same time gets affected by the
CC NSI inputs. We also checked how the NSI parameters can induce a fake
CP-violation ($\delta _{CP}^{\prime }$) effects, when the standard $\delta
_{CP}$ is set to zero in the NSI model. Further, we find constraints on the
CC NSIs at the source and detector for the simulated data of DUNE. 
%

As discussed in section \ref{result} and demonstrated in Fig. \ref{T23CP},
we take two choices to see the impact of CC NSIs on the simultaneous
measurements of $\delta _{CP}$ and $\theta _{23}$. First, we have minimized
over the NSI parameters as shown in Fig. \ref{T23CP}-c and Fig. \ref{T23CP}%
-d, while in the second case we did not marginalize over the NSIs parameters
and this is shown in Fig. \ref{T23CP}-a and Fig. \ref{T23CP}-b. The results
demonstrate that in the absence of CC NSIs, the octant of $\theta _{23}\ $%
can be determined at 3$\sigma $ C.L., while in the presence of the CC NSIs,
the octant cannot be determined at 3$\sigma $ C.L. Similarly, the CC NSIs
also affect the determination of $\delta _{CP}$ and its measurement gets
worsen by approximately 80$\%$ and 50$\%$ in the two cases of different data.

We investigated the effects of NSIs to induce fake CP-violation through the
parameter $\delta _{CP}^{\prime }$. The results of Fig. \ref{DeltaCP} show
that in some cases $\delta _{CP}^{\prime }$ can be large ($\delta
_{CP}^{\prime }\sim 30^{\circ }$), and $\delta _{CP}^{\prime }\sim 0^{\circ }
$ can be excluded by more than $80\%$ C.L. With these results, we can
conclude that the effect of source and detector CC NSIs are important in
determining $\delta _{CP}$ and the CC NSIs can induce large amount of fake
CP-violation.\textit{\ }We showed the near detector of DUNE can help to
constrain the source and detector CC NSIs and distinguish between $\delta
_{CP}^{\prime }$\ and $\delta _{CP}$.

The results on constraining all of the relevant NSI parameters in this study
at the near and far detectors data of DUNE are displayed in Fig. 5. The
related bounds obtained at 90$\%$ C.L. from the one parameter-at-a-time fits
of Fig. \ref{Fig:epsilon} are given in Table I. The results show that DUNE
near detector has a stronger potential to constrain the CC NSIs more
tightly, better than one order of magnitude in comparison to the existing
bounds on the CC NSI parameters at the source and at the detector. The
bounds obtained from the simulated data of the far detector can also
improved by one order of magnitude for some parameters but remain weaker for
the others.

\subsection*{Acknowledgments}

Authors are grateful to Y. Farzan and D. McKay for useful comments and
careful reading of the manuscript. P.B. acknowledges partial support from
the European Unions Horizon 2020 research and innovation programme under the
Marie Sklodowska-Curie grant agreement No. 674896 and No. 690575. A. K. work
has been financially supported by the Sun Yat-Sen University under the
Post-Doctoral Fellowship program and the China Postdoctoral Science
Foundation under the grant \# 74130-41090002.



\begin{thebibliography}{99}
\bibitem{Abe:2011sj} K.~Abe \textit{et al.} [T2K Collaboration], Indication
of Electron Neutrino Appearance from an Accelerator-produced Off-axis Muon
Neutrino Beam, Phys.\ Rev.\ Lett.\  \textbf{107} (2011) 041801,
arXiv:1106.2822. 

\bibitem{DC} Y. Abe \textit{et al}. (Double Chooz Collaboration), Indication
for the disappearance of reactor electron antineutrinos in the Double Chooz
experiment, Phys. Rev. Lett. \textbf{108}, 131801 (2012), , arXiv:1112.6353;
Reactor electron antineutrino disappearance in the Double Chooz experiment,
Phys. Rev. D \textbf{86}, 052008 (2012), arXiv:1207.6632.

\bibitem{RENO} J. K. Ahn \textit{et al}. (RENO Collaboration), Observation
of Reactor Electron Antineutrino Disappearance in the RENO Experiment, Phys.
Rev. Lett. \textbf{108}, 191802 (2012), arXiv: 1204.0626.

\bibitem{DB} F. An \textit{et al}. (Daya Bay Collaboration), Observation of
electron-antineutrino disappearance at Daya Bay, Phys. Rev. Lett. \textbf{108%
}, 171803 (2012), arXiv: 1203.1669; Improved Measurement of Electron
Antineutrino Disappearance at Daya Bay, Chin. Phys. C 37, 011001(2013),
arXiv:1210.6327.

\bibitem{JUNO} Y.-F. Li, Overview of the Jiangmen Underground Neutrino
Detector (JUNO), arXiv:1402.6143v1.

\bibitem{RENO-50} S. Seo, Introduction to RENO-50, in Proceedings of the
International Workshop on RENO-50: Toward Neutrino Mass Hierarchy, Seoul,
2013.

\bibitem{Abe:2014tzr} K.~Abe \textit{et al.} [T2K Collaboration], Neutrino
oscillation physics potential of the T2K experiment, 
PTEP \textbf{2015} (2015) no.4, 043C01, arXiv:1409.7469. 

\bibitem{Adamson:2016xxw} P.~Adamson \textit{et al.} [NOvA Collaboration],
First measurement of muon-neutrino disappearance in NOvA, 
Phys.\ Rev.\ D \textbf{93} (2016) no.5, 051104 arXiv:1601.05037. 

\bibitem{Abe:2015zbg} K.~Abe \textit{et al.} [Hyper-Kamiokande Proto-
Collaboration], Physics potential of a long-baseline neutrino oscillation
experiment using a J-PARC neutrino beam and Hyper-Kamiokande, 
PTEP \textbf{2015} (2015) 053C02, arXiv:1502.05199. 

\bibitem{DUNE} R. Acciarri \textit{et al}. [DUNE Collaboration],
Long-Baseline Neutrino Facility (LBNF) and Deep Underground Neutrino
Experiment (DUNE) Conceptual Design Report Volume 2: The Physics Program for
DUNE at LBNF, arXiv:1512.06148; C. Adams \textit{et al}. [LBNE
Collaboration], The Long-Baseline Neutrino Experiment: Exploring Fundamental
Symmetries of the Universe, arXiv:1307.7335.

\bibitem{Minakata:2013hgk} H.~Minakata and S.~J.~Parke, Correlated,
precision measurements of \$?\_ \{23\} \$ and d using only the electron
neutrino appearance experiments, 
Phys.\ Rev.\ D \textbf{87} (2013) no.11, 113005, arXiv:1303.6178. 

\bibitem{t2khk} K. Abe et al., Letter of Intent: The Hyper-Kamiokande
Experiment --- Detector Design and Physics Potential ---, arXiv:1109.3262.

\bibitem{LBNE} C. Adams et al. (LBNE Collaboration), arXiv:1307.7335

\bibitem{idsf} IDS-NF Collaboration, International Design Study for a
Neutrino Factory, https://www.ids-nf.org/wiki/FrontPage.

\bibitem{essnu} E. Baussan et al. (ESSnuSB Collaboration), Nucl. Phys. B885,
127 (2014).

\bibitem{Coloma:2014kca} P.~Coloma, H.~Minakata and S.~J.~Parke, Interplay
between appearance and disappearance channels for precision measurements of
\$?\_ \{23\} \$ and d, 
Phys.\ Rev.\ D \textbf{90} (2014) 093003, arXiv:1406.2551. 

\bibitem{ANK1} A. N. Khan, D. W. McKay and F. Tahir, Sensitivity of
medium-baseline reactor neutrino mass-hierarchy experiments to nonstandard
interactions, Phys. Rev. D \textbf{88}, 113006 (2013), arXiv:1305.4350.

\bibitem{ANK2} A. N. Khan, D. W. McKay and F. Tahir, Short baseline $%
\overline{\nu }-e$ reactor scattering experiments and nonstandard neutrino
interactions at source and detector Phys.Rev. D \textbf{90}, 053008 (2014),
arXiv:1407.4263.

\bibitem{ANK3} A.~N.~Khan, Global analysis of the source and detector
nonstandard interactions using the short baseline $\nu -e$ and $\overline{%
\nu }-e$ scattering data, 
Phys.\ Rev.\ D \textbf{93} (2016) no.9, 093019, arXiv:1605.09284. 

\bibitem{Girardi:2014kca} I.~Girardi, D.~Meloni and S.~T.~Petcov, The Daya
Bay and T2K results on $\sin ^{2}2\theta _{13}$ and Non-Standard Neutrino
Interactions, 
Nucl.\ Phys.\ B \textbf{886} (2014) 31, arXiv:1405.0416. 

\bibitem{Girardi:2014gna} I.~Girardi and D.~Meloni, Constraining new physics
scenarios in neutrino oscillations from Daya Bay data, 
Phys.\ Rev.\ D \textbf{90} (2014) no.7, 073011, arXiv:1403.5507. 

\bibitem{deGouvea:2015ndi} A.~de Gouvea and K.~J.~Kelly, Non-standard
Neutrino Interactions at DUNE, 
Nucl.\ Phys.\ B \textbf{908} (2016) 318, arXiv:1511.05562. 

\bibitem{Liao:2016hsa} J.~Liao, D.~Marfatia and K.~Whisnant, Degeneracies in
long-baseline neutrino experiments from nonstandard interactions, 
arXiv:1601.00927. 

\bibitem{Coloma:2015kiu} P.~Coloma, Non-Standard Interactions in propagation
at the Deep Underground Neutrino Experiment, 
JHEP \textbf{1603}, 016 (2016), arXiv:1511.06357. 

\bibitem{Coloma:2016gei} P.~Coloma and T.~Schwetz, Generalized mass ordering
degeneracy in neutrino oscillation experiments, 
arXiv:1604.05772. 

\bibitem{Masud:2016gcl} M.~Masud and P.~Mehta, Non-standard interactions and
the resolution of ordering of neutrino masses at DUNE and other long
baseline experiments, 
arXiv:1606.05662. 

\bibitem{Blennow:2016etl} M.~Blennow, S.~Choubey, T.~Ohlsson, D.~Pramanik
and S.~K.~Raut, A combined study of source, detector and matter non-standard
neutrino interactions at DUNE, 
arXiv:1606.08851. 

\bibitem{Bakhti:2016prn} P.~Bakhti and Y.~Farzan, CP-Violation and
Non-Standard Interactions at the MOMENT, 
arXiv:1602.07099. 

\bibitem{Cao:2014bea} J.~Cao \textit{et al.}, Muon-decay medium-baseline
neutrino beam facility, 
Phys.\ Rev.\ ST Accel.\ Beams \textbf{17} (2014) 090101, arXiv:1401.8125. 

\bibitem{ska1} S. K. Agarwalla, P. Bagchi, D. V. Forero and M. Tortola, JHEP
1507 (2015) 060, arXiv:1412.1064.

\bibitem{bkays} D. Dutta, R. Gandhi, B. Kayser, M. Masud and S. Prakash,
JHEP 1611 (2016) 122, arXiv:1607.02152.

\bibitem{ska2} S. K. Agarwalla, S. S. Chatterjee and A. Palazzo, Phys. Lett.
B 762 (2016) 64, arXiv:1607.01745.

\bibitem{valle} F. J. Escrihuela, D. V. Forero, O. G. Miranda, M. Tortola
and J. W. F. Valle, arXiv:1612.07377.

\bibitem{Ohlsson:2012kf} T.~Ohlsson, Status of non-standard neutrino
interactions, 
Rept.\ Prog.\ Phys.\  \textbf{76} (2013) 044201, arXiv:1209.2710. 

\bibitem{Blennow:2015nxa} M.~Blennow, S.~Choubey, T.~Ohlsson and S.~K.~Raut,
Exploring Source and Detector Non-Standard Neutrino Interactions at ESS$\nu $%
SB, 
JHEP \textbf{1509} (2015) 096, arXiv:1507.02868. 
%
%
%

\bibitem{Kopp:2007ne} J.~Kopp, M.~Lindner, T.~Ota and J.~Sato, Non-standard
neutrino interactions in reactor and superbeam experiments, 
Phys.\ Rev.\ D \textbf{77} (2008) 013007, arXiv:0708.0152. 

\bibitem{Kopp:2006wp} J.~Kopp, Efficient numerical diagonalization of
hermitian 3 x 3 matrices, 
Int.\ J.\ Mod.\ Phys.\ C \textbf{19} (2008) 523, arXiv: physics/0610206. 

\bibitem{Adams:2013qkq} C.~Adams \textit{et al.} [LBNE Collaboration], The
Long-Baseline Neutrino Experiment: Exploring Fundamental Symmetries of the
Universe, 
arXiv:1307.7335 [hep-ex]. 

\bibitem{spectrum} http://home.fnal.gov/~ljf26/DUNE2015CDRFluxes/

\bibitem{GonzalezGarcia:2012sz} M.~C.~Gonzalez-Garcia, M.~Maltoni,
J.~Salvado and T.~Schwetz, Global fit to three neutrino mixing: critical
look at present precision, 
JHEP \textbf{1212}, 123 (2012), arXiv:1209.3023. 

\bibitem{Gonzalez-Garcia:2014bfa} M.~C.~Gonzalez-Garcia, M.~Maltoni and
T.~Schwetz, Updated fit to three neutrino mixing: status of leptonic CP
violation, 
JHEP \textbf{1411}, 052 (2014), arXiv:1409.5439. 

\bibitem{Choubey:2016fpi} S.~Choubey and D.~Pramanik, Constraints on Sterile
Neutrino Oscillations using DUNE Near Detector, 
arXiv:1604.04731 [hep-ph]. 

\bibitem{Biggio:2009nt} C.~Biggio, M.~Blennow and E.~Fernandez-Martinez,
General bounds on non-standard neutrino interactions, 
JHEP \textbf{0908}, 090 (2009), arXiv:0907.0097. 

\bibitem{nuno} O.G. Miranda and H. Nunokawa, New J. Phys. 17, no. 9, 095002
(2015), arXiv:1505.06254.

\bibitem{Bakhti:2014pva} P.~Bakhti and Y.~Farzan, Shedding light on LMA-Dark
solar neutrino solution by medium baseline reactor experiments: JUNO and
RENO-50, 
JHEP \textbf{1407}, 064 (2014), arXiv:1403.0744. 

\bibitem{PREM} A. M. Dziewonski and D. L. Anderson, Geophysical aspects of
very long baseline neutrino experiments, Phys. Earth Planet. Interiors 
\textbf{25} (1981) 297; R.~J.~Geller and T.~Hara, 
Nucl.\ Instrum.\ Meth.\ A \textbf{503} (2003) 187, arXiv: hep-ph/0111342. 

\bibitem{Huber:2004ka} P.~Huber, M.~Lindner and W.~Winter, Simulation of
long-baseline neutrino oscillation experiments with GLoBES (General Long
Baseline Experiment Simulator), 
Comput.\ Phys.\ Commun.\  \textbf{167} (2005) 195, arXiv: hep-ph/0407333. 

\bibitem{Huber:2007ji} P.~Huber, J.~Kopp, M.~Lindner, M.~Rolinec and
W.~Winter,New features in the simulation of neutrino oscillation experiments
with GLoBES 3.0: General Long Baseline Experiment Simulator, 
Comput.\ Phys.\ Commun.\  \textbf{177} (2007) 432, arXiv: hep-ph/0701187. 

\bibitem{Messier:1999kj} M.~D.~Messier, Evidence for neutrino mass from
observations of atmospheric neutrinos with Super-Kamiokande, 
UMI-99-23965.

\bibitem{Paschos:2001np} E.~A.~Paschos and J.~Y.~Yu, Neutrino interactions
in oscillation experiments, 
Phys.\ Rev.\ D \textbf{65}, 033002 (2002), arXiv: hep-ph/0107261. 

\bibitem{Acciarri:2016crz} R.~Acciarri \textit{et al.} [DUNE Collaboration],
Long-Baseline Neutrino Facility (LBNF) and Deep Underground Neutrino
Experiment (DUNE) : Volume 1: The LBNF and DUNE Projects, 
arXiv:1601.05471. 

\bibitem{Berryman:2015nua} J.~M.~Berryman, A.~de Gouvea, K.~J.~Kelly and
A.~Kobach, Sterile neutrino at the Deep Underground Neutrino Experiment%
, Phys.\ Rev.\ D \textbf{92} (2015) no.7, 073012, arXiv:1507.03986. 

%
\end{thebibliography}
\end{document}